\documentclass{aa}
\usepackage{graphics}
\begin{document}
\thesaurus{08(08.12.2; 08.12.3);10(10.07.3)}
\title{The Mass Function of NGC\,288}
\author{A. Pasquali\inst{1} \and M.S. Brigas\inst{1,2} \and 
G. De Marchi\inst{1,3,4}} 
\institute{ESO/ST-ECF, Karl-Schwarzschild-Strasse 2, D-85748 Garching
bei M\"unchen, Germany \and Osservatorio Astronomico di Cagliari,
Strada 54, Poggio dei Pini, 09012 Capoterra, Cagliari, Italy \and
STScI, 3700 San Martin Drive, Baltimore, MD 21218, USA \and Affiliated with 
the Astrophysics Division, Space Science Department of ESA}
\offprints{A. Pasquali}
\mail{Anna.Pasquali@eso.org}
\date{}
\maketitle
 
\begin{abstract}
We present NICMOS NIC3 observations of the Galactic globular cluster
NGC\,288, taken South - East at 2.4 times the cluster's half-light
radius in the $J$ and $H$ bands. We have detected the cluster main
sequence down to $J \simeq 25$  and  $H \simeq 24$. The corresponding
luminosity function covers the range $3 < M_H < 9$ and peaks at $M_H
\simeq 6.8$.  The theoretical tracks of Baraffe et al. (1997) at
$[Fe/H] = -1.3$ give a mass function which is best fitted by a
log-normal distribution with characteristic mass of $m_c =
0.42$\,M$_{\odot}$ and a standard deviation $\sigma = 0.35$. This
result is fully consistent with the global mass function derived by
Paresce \& De Marchi (2000) for a sample of 12 globular clusters with
very different dynamical histories, thus confirming that near the
cluster's half-light radius the  mass function appears ``undistorted''
by evaporation or tidal interactions with the Galaxy and should then
reflect the initial mass function.

\end{abstract}

\section{Introduction}

The Galactic globular cluster NGC\,288 is located about one degree away
from the South Galactic Pole, in a region of negligible interstellar
extinction, $E(B-V) = 0.04$ (Alcaino et al. 1997), and very low
contamination by field stars. The cluster is at a distance modulus of
$(m-M)_V = 14.7$ and is relatively metal-poor with $[Fe/H] = -1.3$ 
(Alcaino et al. 1997).  Its stellar population is characterized by the
presence of blue stragglers and by an anomalous horizontal branch (HB)
which is almost entirely composed of stars blue-ward of the instability
strip (Alcaino et al. 1997). Such a blue HB may well be induced by the
second parameter, although the study of Buonanno et al. (1984) ruled
out that the HB anomaly observed in NGC\,288 is due to any peculiarity
in the cluster's He content, metallicity or age. Buonanno et al. (1984)
estimated an age of $16 \pm 3$\, Gyr from the $\Delta m_{(TO-HB)}$
method, whereas Alcaino et al. (1997) have derived $14 \pm 2$ Gyr from
isochrone fitting techniques.
   
NGC\,288 has been observed at infrared wavelengths only by Davidge \&
Harris (1997), who have acquired 3\,min exposures in $J$ and 3\,min
exposures in $K$ of a central field and 48\,min images both in $J$ and
$K$ of a region at $140^\prime\prime$ West and South of the cluster
core. In this way, Davidge \& Harris have resolved the red super-giant
branch of the cluster down to the upper main sequence (MS) at $K < 20$ and
$0 < J-K < 1$.  Their isochrone fitting has been performed with the
oxygen-enriched (at $[Fe/H] = -1.26$) tracks of Bergbusch \& Vandenberg
(1992) and has resulted into an age in excess of 16\,Gyr.
   
In this paper, we present the first NICMOS observations of a field in
the outskirts of NGC\,288 taken in the $J$ and $H$ bands. The data are
described in Section\,2 and the cluster's colour-magnitude diagramme
(CMD) and mass function (MF) are presented in Section\,3. Discussion
and conclusions follow in Section\,4.

\section{Observations and data reduction}

NGC\,288 has been observed with the NIC3 camera of the NICMOS
instrument on board the HST during the parallel observations campaign
(GO 7811). The observed region is located $2.4$ times the cluster's
half-light radius ($r_{hl}=2\farcm25$; Djorgovski 1993) away from the
centre, or $1\farcm6$\,E and $5\farcm1$\,S.  Six images of the same
field are available, both in the $J$ and $H$ bands (NIC3 filters F110W
and F160W, centered at $1.1\,\mu$m and $1.6\,\mu$m, respectively) for a
total exposure time of 45\,min in $J$ and 47\,min in $H$.  The detailed
log of the observations is given in Table\,1.

\begin{table}
\caption[]{Log of the observations}
\vspace{0.3cm}
\begin{tabular}{l c c}
\hline
Image     & Filter & Exposure time (s)\\
\hline
N4EZ07CWQ & F110W & 575.94\\
N4EZ07CXQ & F160W & 575.94\\
N4EZ07D1Q & F110W & 575.94\\
N4EZ07D3Q & F160W & 575.94\\
N4EZ08D8Q & F110W & 575.94\\
N4EZ08D9Q & F160W & 575.94\\
N4EZ08DDQ & F110W & 575.94\\
N4EZ08DFQ & F160W & 575.94\\
N4F001CMQ & F110W & 191.96\\
N4F001CNQ & F160W & 255.96\\
N4F001CPQ & F110W & 191.96\\
N4F001CQQ & F160W & 255.96\\
\hline
\end{tabular}
\end{table}

The images have been reduced using the NICMOS standard calibration
pipeline: they have been first processed with CALNICA for bias
subtraction, dark-count correction and flat-fielding. They have then
been associated and combined with CALNICB, to remove cosmic rays and to
increase the signal-to-noise ratio. Photometry on the images has been
performed with the DAOPHOT package. Stars have been identified with
DAOFIND, by setting the detection threshold at $10\, \sigma$ above the
background. We have traced the radial profile of each identified object
and discarded those which showed a full width at half-maximum (FWHM)
larger than $2\farcs5$, being the typical FWHM of a point source
$1\farcs5$ in our frames. This procedure has produced a sample of 75
stars. We have also detected 15 extended objects, whose fundamental
parameters (flux, position and shape parameters) have been measured
with S-Extractor.
   
The field not being crowded, stellar count-rates have been measured in
fixed apertures of 5\,pixel radius (equivalent to 1$''$), and the
corresponding background values have been determined in a fixed annulus
with a radius of 7\,pixel and a width of 3\,pixel.  After background
subtraction, stellar count-rates have been corrected for the NIC3
intra-pixel sensitivity, using the equations computed by Storrs et al.
(1999, cf. Table 2) in the case of out-of-focus campaign data, and for
the camera being out of focus, making use of TinyTim (Krist \& Hook,
1999) which simulates the PSF of the NIC3 camera with the precise
optics settings corresponding to a specific filter and to a specific
observation date. We have used TinyTim to compute two PSFs for each
frame, one for our observation date (November 1997) and one for January,
15, 1998 when NIC3 was in-focus (in-focus campaigns were carried out in
1998 January and June). We have calculated the encircled energy for a
5\,pixel aperture for each PSF and used the flux ratio in-focus to
out-of-focus to correct our measured count-rates.  Finally, we have
multiplied the sample count-rates by a factor of $1.075$ to reconduct
them to the values measured in a nominal infinite aperture.
   
The corrected count-rates ($c$) have been converted into the VEGAMAG
photometric system by means of the relation:
   
$$
m = -2.5\,\,log \left(\frac{c\,U}{Z}\right)
$$
   
where $U$ is the count-rate/flux conversion factor (known as {\it
inverse sensitivity}) and $Z$ is the flux of a zero-magnitude star in
the VEGAMAG system, provided for all NICMOS filters and VEGAMAG bands
by the HST Data Handbook and by the NICMOS Photometry Update (cf.
http://www.stsci.edu/instruments/nicmos).
    
The main source of error in our photometry is due to the intra-pixel
sensitivity effect, that causes up to 30\% flux variations among
individual images. With the correction of Storrs et al. (1999) the
spread in photometry is reduced to 0.1 mag in J and 0.09 mag in H.  We,
therefore, assume an uncertainty of 0.1 mag in both the $J$ and $H$
bands.

\section{Results}

\subsection{Photometric completeness}

The dereddened ($E(B-V) = 0.04$; Alcaino et al. 1997) $J$ and $H$
magnitude distributions are shown in Figure\,1 for the samples of stars
and galaxies identified in the observed NGC\,288 field. Stars populate
the $\sim 18 < J < \sim25$ and $\sim18 < H < \sim24$ intervals, while
galaxies distribute in the $\sim17 < J < \sim24$ and $\sim19 < H <
\sim24$ ranges.  The fact that stars and galaxies as faint as $J$ or $H
\simeq 24$ have been detected indicates that the cluster field does not
contain objects fainter than the $24$th mag and, thence, that our star
sample is complete. Completeness is also secured by the very low level
of field crowding, which allows full object detection by eye.

\begin{figure*}
\resizebox{12cm}{!}{\includegraphics{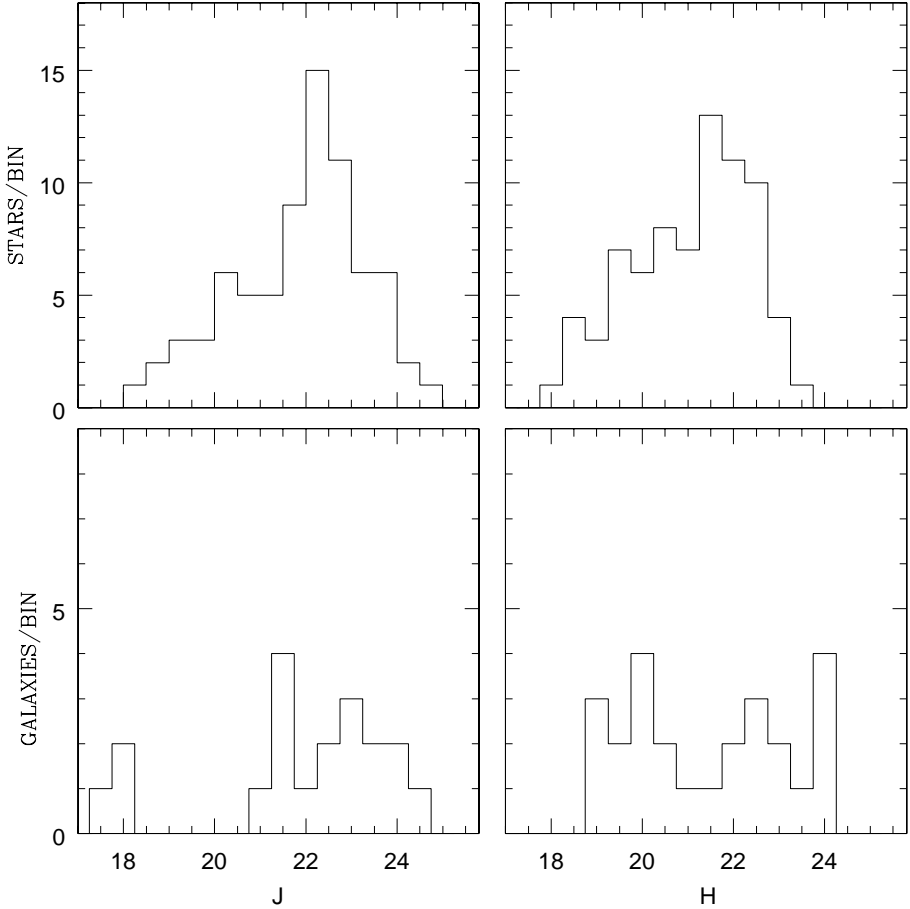}}
\caption{Magnitude distributions of the star and galaxy samples
detected in the observed field}
\end{figure*}

\subsection{The Colour-Magnitude Diagramme}

The dereddened CMD of the stars in our sample is plotted in Figure\,2.
It extends over the $H$-band magnitude interval $18 - 24$ and the $0.1
< J-H < 1.4$ in colour.

\begin{figure}
\resizebox{7cm}{!}{\includegraphics{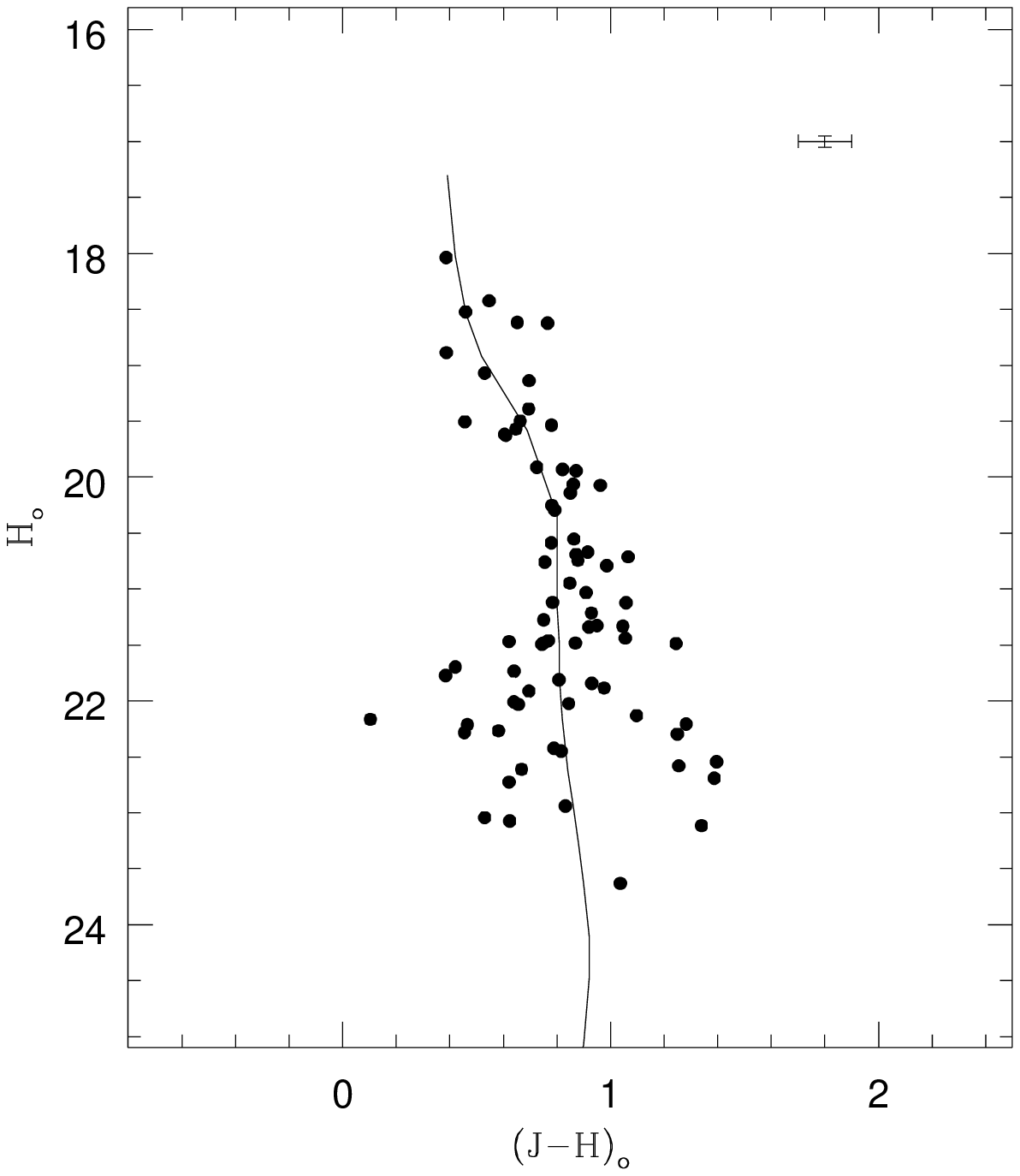}}
\caption{Dereddened colour-magnitude diagram. The theoretical track of
Baraffe et al. (1997) computed for $[Fe/H] = -1.3$ has been superposed
to the observed distribution. Uncertainties amount to $0.1$ mag and
$0.2$ mag on $H$ magnitudes and on $J-H$ colours, respectively.}
\end{figure}

We have superposed on the observed distribution the expected $H$-band
magnitudes and $J-H$ colours as computed for low mass stars at $[Fe/H]
= -1.3$ by Baraffe et al. 1997 (solid line). The theoretical track is
here scaled by the NGC\,288 distance modulus $(m-M)_0 = 14.7$. The IR
colours provided by Baraffe et al. have been computed adopting the most
recent non-grey model atmospheres and an improved equation of state for
low mass stars. The quite good agreement between the observed and the
theoretical distributions (within an observational uncertainty of
$0.2$\,mag in colour) confirms that we have indeed detected the lower
end of the MS of NGC\,288.
   
Particularly evident in Figure\,2 is the change of slope occurring at
at $H \simeq 20$ and $J-H = 0.8$. This is due to the atmospheric
opacity being mostly produced by H$_2$ molecules for stellar masses
lower than $0.5$\,M$_\odot$.

\subsection{The luminosity and mass functions}

In spite of the limited number of stars in our sample, we have been
able to trace the luminosity function (LF) for the observed cluster
field, as shown in Figure\,3, where the number of stars observed in
each $0.5$\,mag bin is plotted as a function of the $H$-band magnitude.
The LF displays the same general features found by Paresce \& De Marchi
(2000) for a sample of twelve Galactic globular clusters: a peak at
$M_H \simeq 6.8$ (corresponding to $M_I \simeq 8.2$) followed by a
rapid decrease towards fainter luminosities.

\begin{figure}
\resizebox{7cm}{!}{\includegraphics{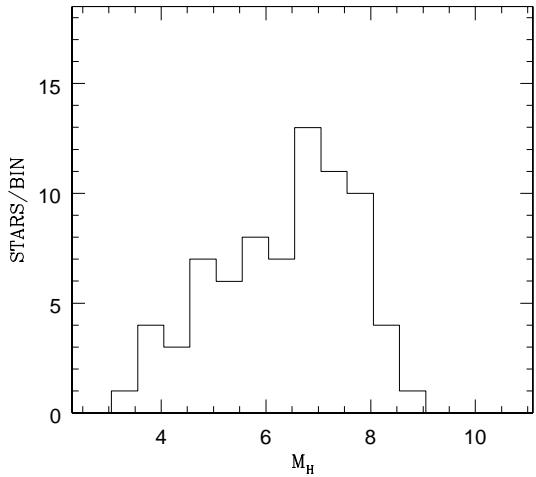}}
\caption{The luminosity function of the field in the H band.
A distance modulus (m-M)$_0=14.7$ has been adopted}
\end{figure}

We have used the mass-luminosity relation corresponding to the
theoretical track plotted in Figure\,2 to convert the LF into the cluster's
MF. Since the observed LF is the product between the MF
and the derivative of the mass-luminosity relation, we have first
adopted a specific MF, subsequently derived the mass-luminosity of
Baraffe et al. and computed their product. The latter has finally been
compared with the observed LF of Figure\,3 until a good fit was found.
As our first attempt, we have assumed a power-law mass function in
the form of $dN/dlog(m) \propto m^{-x}$ (using this notation, Salpeter's
IMF would have $x=1.35$). We have found, however, that no single
power-law mass function can fit the LF of the observed NGC\,288 field.
Indeed, an index $x = -0.1$ can well reproduce the bright portion of
the observed LF at $M_H < 6.8$, but it overestimates the number of
stars fainter than the LF peak (cf. top panel of Figure\,4). On the
other hand, an index $x = -1.1$ properly fits the faint section of the
LF at $M_H >  6.8$, but it predicts a number of bright stars a factor
of $\sim$2 larger than observed (cf.  bottom panel of Figure\,4).

\begin{figure}
\resizebox{7cm}{!}{\includegraphics{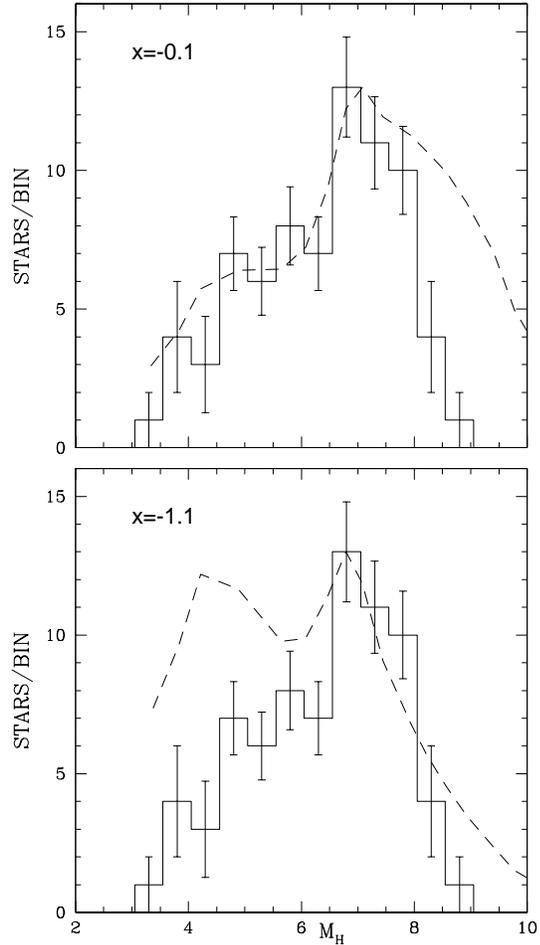}}
\caption{Two power-law mass functions, $dN/dlog(m) \propto m^{-x}$
have been fitted to reproduce the luminosity function observed for
NGC\,288.  Top panel: an index $x = -0.1$ well matches the bright tail
of the LF at $M_H < 6.8$ but fails to reproduce the observed number of
fainter stars. Bottom panel: an index $x = -1.1$ fits well the faint
tail of the LF at $M_H > 6.8$, yet it overestimates the number of
bright stars by nearly a factor of 2}
\end{figure}

As an alternative to a power-law, Paresce \& De Marchi (2000) have
shown that a log-normal distribution peaked at $\sim 0.35$\,M$_\odot$
gives a good fit to the MF of globular clusters for masses smaller than
$\sim 0.8$\,M$_\odot$ (i.e. for stars that are still on their MS). Our
data show that the LF obtained for MS stars in NGC\,288 is fully
compatible with a log-normal distribution of the form:

$$
ln\left(\frac{dN}{dlog(m)}\right) = A -
\left[\frac{log(m/m_c)}{\sqrt{2}\sigma}\right]^2
$$
   
where A is a normalization constant, provided that the characteristic
mass takes on the value of $m_c = 0.42$ and the  standard deviation is
$\sigma = 0.35$ (see Figure\,5).

\begin{figure}
\resizebox{7cm}{!}{\includegraphics{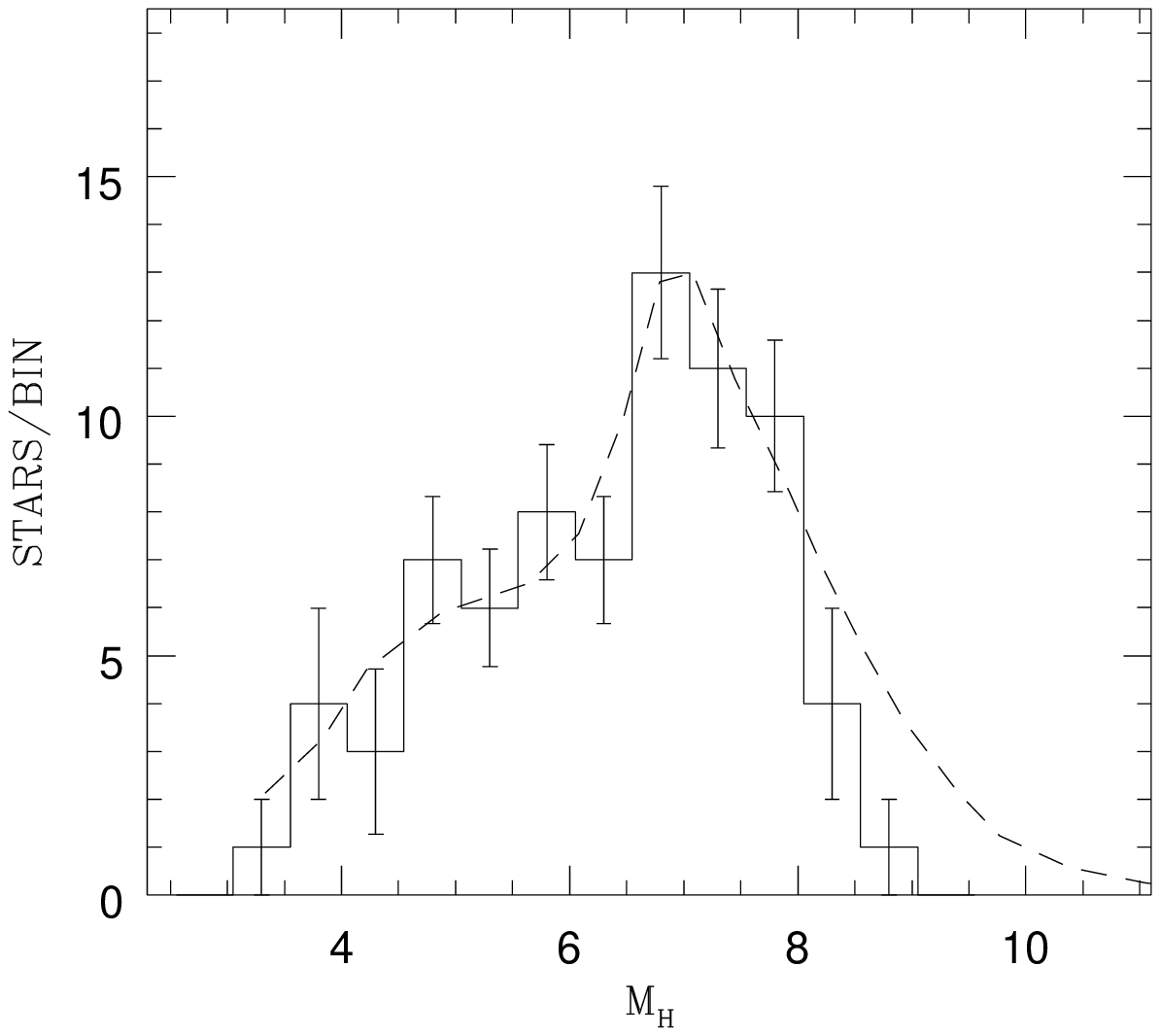}}
\caption{The best fit to the observed luminosity function is given by
the log-normal distribution ln(dN/dlog(m)) = A -
[log(m/m$_c)/{\sqrt{2}\sigma}]^2$ where m$_c = 0.42$\,M$_{\odot}$ and
$\sigma = 0.35$}
\end{figure}

We note here, as already pointed out by Paresce \& De Marchi (2000),
that the fact that a log-normal distribution is a viable form for the
MF of a cluster for stars with mass $m < 0.8$\,M$_\odot$ does not imply
that the same functional form is appropriate at higher masses as well,
where a power-law distribution seems more appropriate (see Elmegreen
1999).

\section{Discussion and conclusions}

Comparing the best fitting LF of Figure\,5 with the results published
by Paresce \& De Marchi (2000) indicates that the MF of NGC\,288 is
fully compatible with the global function found by these authors for a
sample of twelve Galactic clusters spanning a wide range of
metallicity, different distances from the Galactic centre and plane,
and structural parameters (core and half-light radius and concentration
ratio).
   
NGC\,288 is characterized by a highly elliptical orbit which makes
disruption by tidal shocking quite effective and more important than
disruption by internal two-body relaxation, according to the
calculations of Dinescu et al. (1999). In particular, Gnedin \&
Ostriker (1997) have derived a disruption time of $\sim 1$\,Gyr for
NGC\,288,  which suggests that this cluster might have experienced a
strong interaction with the Galactic tidal field during its lifetime,
such that it could be totally dissolved within the next Gyr or so.

Using the log-normal MF determined so far, we have computed the index
$\Delta\,\log\,N$ defined as the logarithmic ratio between the
number of stars with mass $m=m_c= 0.42$\,M$_{\odot}$ and the number of
stars with $m = 0.7$\,M$_{\odot}$. Paresce \& De Marchi (2000) have
defined such a parameter in order to quantitatively describe the shape
of the MF and correlate it with the properties of the cluster (see
their paper for more details). In Figure\,6, we have plotted the value
of $\Delta\,\log\,N$ for NGC\,288 along with those of the clusters
studied by Paresce \& De Marchi as a function of their expected time to
disruption.

We note here that Paresce \& De Marchi have measured $\Delta\,\log\,N$
on the global MF of the cluster, i.e. the MF after correction of the
effects of internal dynamical evolution (mass segregation). The global
MF has been shown by De Marchi, Paresce \& Pulone (2000) to closely
approach the local MF, provided the latter is measured near the
cluster's half-light radius ($r_{hl}$). Our data, however, have been
taken at $\sim 2.4\,r_{hl}$, so that a correction would be required to
our value of $\Delta\,\log\,N$ for NGC\,288. Unfortunately, we do not
have enough data (surface brightness and radial velocity profiles and
more than one LF at different radial locations) to investigate the
dynamical structure of the cluster and correct for the effects of mass
segregation. On the other hand, as shown by De Marchi et al. (2000), at
$\sim 2.4 \, r_{hl}$ the MF is expected to be steeper than the global
MF in the mass range of interest here ($0.4 - 0.8$\,M$_\odot$) and, as
such, the value of $\Delta\,\log\,N$ is to be considered here an upper
limit.

\begin{figure}
\resizebox{9cm}{!}{\includegraphics{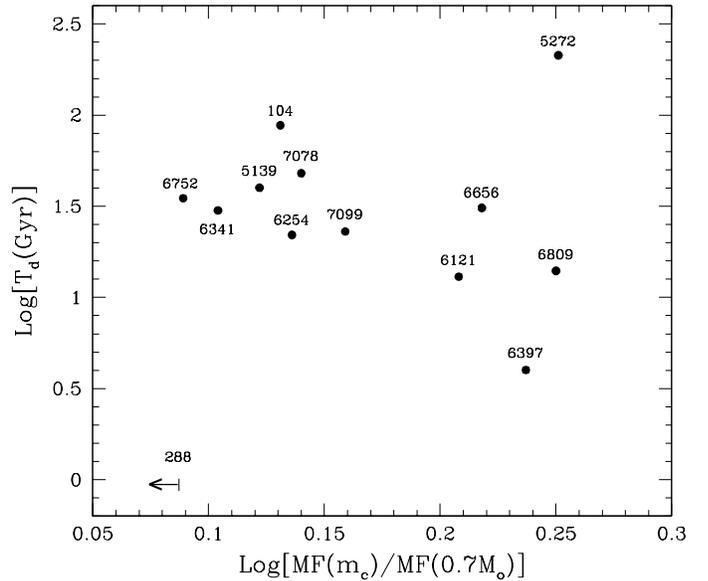}}
\caption{The plane (Disruption Time) vs.
log[MF($m_c$)/MF(0.7M$_{\odot}$)] from Paresce \& De Marchi (2000)
where NGC\,288 has been added. The position  of NGC\,288 (upper limit
in $\Delta\,\log\,N$) confirms the scattered nature of the distribution
so that no correlation between dynamical evolution and mass function
may be drawn}
\end{figure}

Thus, the position of NGC\,288 at the opposite corner of NGC 5272
confirms the scattered nature of the distribution in Figure\,6, hence
the absence of any correlation between dynamical evolution and the MF
observed near the clusters half-light radius as pointed out by Paresce
\& De Marchi. Indeed, compared to the bulk of globular clusters with a
mean disruption time $T_d \simeq 32$\,Gyr, NGC\,288 appears to have
suffered the highest degree of erosion from the Galaxy. Yet, its MF is
nearly identical to what observed for the cluster sample of Paresce \&
De Marchi (2000). So, although it has not been determined precisely at
the $r_{hl}$, it is most likely that the observed MF closely reflect
the cluster's initial mass function (IMF).  An example of the opposite
case is offered by NGC\,6712, a cluster with a dynamical history not to
different from that of NGC\,288, yet with a very different MF.  Its MF,
measured by De Marchi et al. (1999) at $1.7 \, r_{hl}$ peaks at
$0.75$\, M$_{\odot}$ and slowly drops down to 0.3 M$_{\odot}$, while
the average MF of all the clusters in Figure\,6 is always increasing in
the mass range $0.75 - 0.3$\,M$_{\odot}$. The lack of low-mass stars
(less massive than 0.75 M$_{\odot}$) in NGC\,6712 has been attributed by
De Marchi et al. (1999) and by Takahashi \& Portegies Zwart (2000) to 
the effect of stripping by tidal interactions with the Galactic bulge.
   
According to Pryor \& Meylan (1993), NGC\,288 and NGC\,6712 have a
concentration ratio of $0.96$ and $0.90$, respectively, so that they
have so far experienced a similar degree of internal two-body
relaxation.  NGC\,6712 is twice more massive but a factor of 4 closer
to the Galactic centre. The cluster orbits have the same ellipticity
($0.75$) but different orbital periods, namely $230 \times 10^6$ yr
and $130 \times 10^6$ yr for NGC\,288 and NGC\,6712, respectively.
Given an apogalactic distance $R_a$ of 11\, kpc ($R_p = 1.8$\, kpc) and
an observed Galactocentric distance $R_{GC} = 11.1$\,kpc (Dinescu et
al.  1999), NGC\,288 is now at its apogalactic point. On the contrary,
NGC\,6712, with $R_a = 6$\,kpc ($R_p = 0.9$\,kpc) and $R_{GC} =
3.5$\,Kpc, is only half-way between its perigalactic and apogalactic
points, so that it has crossed the Galactic bulge more recently than
NGC\,288. Based on the difference in the orbital phase alone, one
might speculate that NGC\,288 has been able to thermalize its mass
distribution after the bulge shock while NGC\,6712 is still suffering
from it. This hypothesis is, however, not applicable as Gnedin \&
Ostriker (1997) give a relaxation time at half-light radius of
$1.4$\,Gyr and $0.7$\,Gry for NGC\,288 and NGC\,6712, respectively,
which is a factor of 5 to 6 larger than the clusters orbital period.
Hence, both clusters have still to reach internal relaxation after the
last bulge shocking.

Without more data on the LF of these clusters much further out in their
periphery, it is not possible to fully understand the differences
between their MF. It is, however, plausible that NGC\,6712, given its
orbital parameters, plunges more deeply than NGC\,288 into the bulge so
that the low-mass stars depletion observed in its LF is essentially due
to stripping by bulge crossing or that the severe modification to its
MF has been impressed by encounters with molecular clouds existing in
the spiral arms of the Milky Way where this cluster currently sits.

\begin{acknowledgements}
We would like to thank Isabelle Baraffe for computing the theoretical
tracks at $[Fe/H] = -1.3$ in the NICMOS $J$ and $H$ bands and Oleg
Gnedin for helpful discussions. We also thank an anonymous referee for
valuable comments and suggestions that have considerably strengthened
the presentation of this work.  MSB acknowledges support from the
Osservatorio Astronomico di Cagliari and from the Director General's 
Discretionary Fund at ESO.
\end{acknowledgements}

\end{document}